\documentclass[twocolumn,aps,prl,superscriptaddress,floatfix]{revtex4-1}
\usepackage{graphicx}
\usepackage{amsmath}

\begin{document}
\title{Stable and Metastable Kinetic Ferromagnetism on a Ring}

\author{Ilya Ivantsov}
\affiliation{Bogoliubov Laboratory of Theoretical Physics, Joint
Institute for Nuclear Research, Dubna, Russia}
\author{Hernan B. Xavier}
\affiliation{International Institute of Physics - UFRN,
Department of Experimental and Theoretical Physics - UFRN, Natal, Brazil}
\author{Alvaro Ferraz}
\affiliation{International Institute of Physics - UFRN,
Department of Experimental and Theoretical Physics - UFRN, Natal, Brazil}
\author{Evgenii Kochetov}
\affiliation{Bogoliubov Laboratory of Theoretical Physics, Joint
Institute for Nuclear Research, Dubna, Russia}

\begin{abstract}
Performing an exact diagonalization of the effective spin problem, a ferromagnetic ground state of kinetic origin is shown to emerge in a system of $N$ strongly correlated electrons on a $L$-site ring ($L > N$).
This phenomenon is brought about by the \textit{quantum necklace} statistics originated from the no double occupancy constraint leading to a fractional shifted electron momentum quantization.
As a consequence of such special energy level distribution, the kinetic ferromagnetism is stable only for $N=3$.
For odd $N>3$ the fully polarized FM state energy is only a local minimum but it is protected by a finite energy barrier that inhibits one spin-flip processes.
The metastable ferromagnetic state survives perturbations of small magnitude opening up a possibility of being experimentally observed by an appropriate tuning of the interdot tunneling amplitudes in currently available quantum dot arrays.
\end{abstract}

\maketitle


Kinetic (itinerant) ferromagnetism (FM) had been just a theoretical prediction during the last 50 years before it was recently
experimentally observed in quantum dot arrays \cite{Dehollain2019}.
According to numerical calculations \cite{Takahashi1982,Buterakos2019,Wang2019}, three strongly correlated electrons in the $2\times 2$ plaquette do form such FM ground state of kinetic origin.
The key feature behind this phenomenon is the emergence of the FM ordering
with a strong electron correlation encoded into the no double electron occupancy constraint and no 
finite-range exchange spin-spin interaction.

There are only a few rigorous theoretical results for kinetic FM, with the Nagaoka's ferromagnetism
being the most important one \cite{Nagaoka1965, Nagaoka1966}.
Nagaoka's theorem predicts the FM ground state in $2D$ systems with one hole in a half-filled band, provided certain conditions hold.
However, creating precisely one hole in a thermodynamic system is obviously an impossible constraint.
Moreover, there is also no consensus about the presence\cite{Becca2001,Liu2012} or not \cite{Putikka1992,Ivantsov2017} of
the stable kinetic FM at a non-zero finite hole concentration in the thermodynamic limit, or even about 
on the role of the electron-electron interaction in this phenomenon \cite{Kornilovitch2014}.


%
Nagaoka's proof implies that the so-called connectivity condition holds: any two spin configurations can be transformed into one another by an appropriate electron hopping process.
This is the case for a bipartite $2D$ and for regular lattices of higher dimension.
However it breaks down for a $1D$ ring.
In this case Nagaoka's theorem does not apply.
At first sight it may seem that the $1D$  solution is trivial  since in a chain with open boundary conditions (BCs) the spectrum
is completely degenerate with respect to its spin content. The existence of boundaries as well as the impossibility of exchanging
the spin relative ordering prevents the projected electrons to access different spin configurations. Because of this the projected electrons behave
essentially as spinless fermions. As a matter of fact, the Lieb-Mattis theorem tells us that
in this case the ground state is realized as the lowest total spin state\cite{Lieb1962}.

In contrast, for periodic BCs the situation changes drastically.
The underlying configuration space is now a closed loop - a ring - with $L$ sites and $N\le L$ projected electrons.
As soon as an electron crosses the boundary the corresponding spin configuration undergoes a cyclic permutation.
In this way the emergent spin configurations are decomposed into
disconnected parts -- the so called quantum "necklaces" -- that cannot be transformed  into one another  by permutations\cite{Xavier2019}.
The importance of the closed loops is well-known, and may lead to some rather unexpected non-trivial results, such as an emergence of the kinetic AF order on frustrated lattices \cite{Haerter2005}.


The aim of the present article is twofold. First we derive the exact conditions under which the kinetic FM emerges on an $L$ - site ring accommodating $N$ electrons.
Second, we discuss in what way our findings can be used in experimental setups to observe the kinetically driven FM states.

\paragraph*{Exact diagonalization of 1D ring.}

Our consideration is based on the $t-J$ model of strongly correlated electrons:
\begin{equation}
H=-\sum_{ij\sigma}t_{ij}\tilde{c}^{\dagger}_{i\sigma}\tilde{c}_{j\sigma}+J\sum_{\langle ij\rangle}\mathbf{S}_i\cdot \mathbf{S}_j.
\label{Eq1}
\end{equation}
Here $\tilde{c}^{\dagger}_{i\sigma}=(1-n_{i,-\sigma})c^{\dagger}_{i\sigma}$ is a constrained electron creation operator
on a site $i$ with the spin projection $\sigma=\uparrow,\downarrow$. The symmetric matrix
$t_{ij}$
represents the hopping amplitude with $t>0$ between the nearest neighbour sites  and zero otherwise. The $\mathbf{S}_i$ stands
for the electron spin operator.
This model captures both the antiferromagnetic (AF) ordering instability caused by
the induced spin exchanges between nearly localized electrons $(\sim J)$
as well as the FM itinerant magnetism due to the possible gain in the particle kinetic energy.

The model (\ref{Eq1}) becomes exactly solvable in the limit of $J\to 0$ in a $1D$ lattice. In this case, it exhibits
a spin charge separation in the following way: the $N$-electron wave function of the $L$-site ring is decomposed as $|\psi\rangle=|\psi_r\rangle|\psi_S\rangle=|r_1,r_2,...,r_N\rangle|s^z_1,s^z_2,...s^z_N\rangle$ where $r_i=1,..,L$ is the
and $s^z_i=\pm\frac{1}{2}$ is the spin projection of the $i$-th electron.
Due to the no double occupancy constraint and the presence of only  nearest neighbours hopping the spin configurations
are not affected by bulk hopping and they simply undergo a cyclic permutation under crossboundary hopping.
In this representation the Hamiltonian takes the form:
\begin{equation}
H=-t\sum_{i=1}^{L-1}c^{\dagger}_{i}c_{i+1}-t~c^{\dagger}_{L}c_{1}\hat{P}+\mathrm{h.c.},
\label{Eq2}
\end{equation}
where $c_i$ is the spinless fermion annihilation operator on the site $i$ that acts only on the spatial part of wave function $|\psi_r\rangle$. The $\hat{P}$ is the cyclic spin permutation operator which affects only the spin part $|\psi_S\rangle$
\begin{equation}
\hat{P}\cdot |s^z_1,s^z_2,...,s^z_N\rangle=|s^z_N,s^z_1,...,s^z_{N-1}\rangle.
\end{equation}

Since the spin part of the total wave function is affected only by the cyclic permutation operator $\hat{P}$  we can diagonalize the spin part of Hamiltonian Eq.(\ref{Eq2}) separately from the spatial part.
It should be noted that $\hat{P}$ is the block-diagonal matrix where each block corresponds to the set of configurations which can be obtained from each other by a cyclic permutation.
Indeed, the $\hat{P}$ operator connects the $|\uparrow\downarrow\uparrow\downarrow\rangle$ with the $|\downarrow\uparrow\downarrow\uparrow\rangle$ but not with $|\uparrow\uparrow\downarrow\downarrow\rangle$.
Thereby the configurations connected by the cyclic permutation are inside one single spin block.
The size of each block (the number of connected configurations inside the $j$-th block) is $N_j$, where $j$ enumerates
all possible disconnected spin blocks.
In general the $N_j=N/\phi_j$ where $\phi_j$ is an integer depending on the symmetry of the configuration.
In this way we can calculate all eigenvalues and eigenstates of $\hat{P}$:
\begin{equation}
\begin{split}
\lambda_j(p)&=e^{2\pi i p/N_j}\\
|\psi_{j}(p)\rangle&=\frac{1}{\sqrt{N_j}}\sum_{q=0}^{N_j-1}e^{2\pi i pq/N_j}\cdot\hat{P}^{q}~|\tilde{\psi}_{j}\rangle,
\end{split}
\end{equation}
where $|\tilde{\psi}_{j}\rangle$ is any configuration from the $j$-th block.
The values $p=0..N_j-1$ enumerate the eigenvalues within a given
fixed $j$-th block.
Physically, the $p$ corresponds to the spin wave momentum induced by the cyclic permutation.


In this basis, the Hamiltonian Eq.(\ref{Eq2}) becomes block diagonal.
The index $j$ together with all available values of $p$ may be considered as the spin quantum numbers.
In each $j$-th block we have
\begin{equation}
H_j(p)=-t\sum_{i=1}^{L-1}c^{\dagger}_{i}c_{i+1}-t~c^{\dagger}_{L}c_{1}e^{2\pi i p/N_j}+\mathrm{h.c.}.
\label{Eq3}
\end{equation}
While the indexes $j$ and $p$ define all possible spin configurations, $H_j(p)$ describes the corresponding spatial distribution of the electrons.
It is easy to see that it is essentially a tight-binding Hamiltonian of spinless electrons on a ring  with an
effective "magnetic" flux $2\pi p/N_j$ threading through the ring.
Note that the solution of Eq.(\ref{Eq3}) does not depend on a specific form of the spin wave function,
rather it is determined by the values of $p$.

\begin{figure}
\includegraphics[width=1\linewidth]{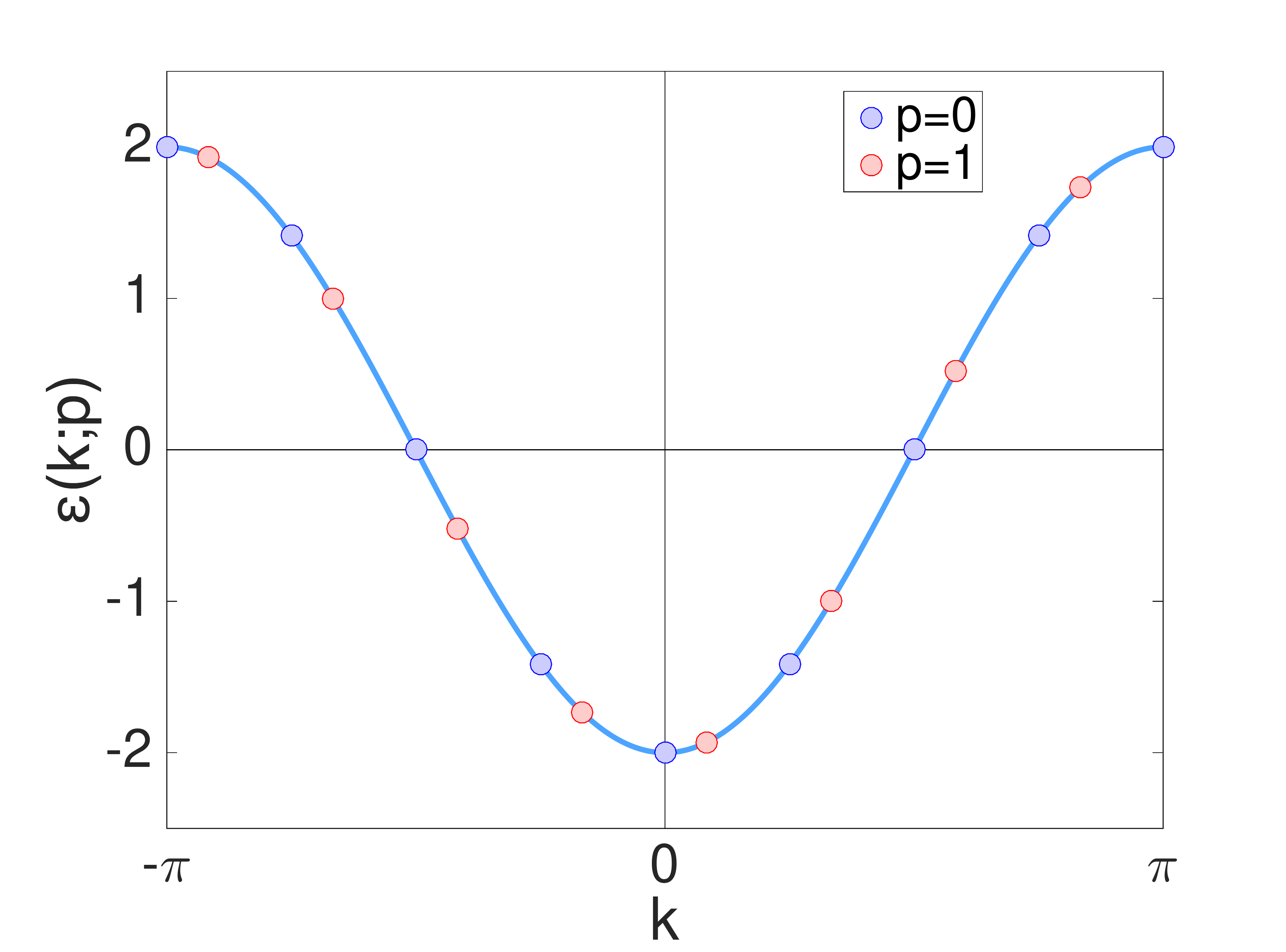}
\caption{Momentum quantization at various values of $p$ for the $L=8$ ring at $N=3$.}
\label{Fig1}
\end{figure}

The one-electron solution is therefore a tight-binding electron wave function,
\begin{equation}
\begin{split}
|\psi(k,p)\rangle&=\frac{1}{\sqrt{L}}\sum_{x}e^{ix\frac{2\pi}{L}(k+\frac{p}{N_j})}c_{x}^{\dagger}|0\rangle\\
\varepsilon(k;p)&=-2t\cos(\frac{2\pi}{L}(k+\frac{p}{N_j})),
\end{split}
\end{equation}
with the electron momentum shifted by $2\pi p/(LN_j)$ (see Fig.\ref{Fig1}).
The simplest example of such a solution is a single electron wave function in the absence of correlations.
In this case all $N_j=1$ and $p=0$ for any $j$. As a result, the wave function reduces to that of a free electron as it should be.

For the many electron case the many body wave function is the Slater determinant of the single electron wave functions with the momenta $\mathbf{k}:=(k_1,k_2,...,k_N)$ and with a fixed value of $p$ which is denoted as $|\Psi_{el}(\mathbf{k},p)\rangle$.
As a result, all eigenstates and eigenvalues of Hamiltonian (\ref{Eq2}) can be written as:
\begin{equation}
\begin{split}
|\Psi(\mathbf{k};j,p)\rangle&=|\Psi_{el}(\mathbf{k},p)\rangle\cdot|\psi_j(p)\rangle,\\
E(\mathbf{k};j,p)&=\sum_{i=1}^{N}\varepsilon(k_i;p),
\end{split}
\label{eig}
\end{equation}
where $p=0..N_j-1$ for each spin block $j$.
In this way the eigenenvalues are determined by both the values of electron momenta  $\mathbf{k}$ and the value of the spin wave momentum $p$.
The correlations come into play through the $p/N_j$ corrections to the electron momentum.
This is a collective phenomena since the $N_j$ is proportional to a total number of electrons $N$.
For a given spin configuration the available values of $p$ determine the eigenenergies.

Explicitly these wave functions can be written in the following way:
The fully polarized state $|\psi_{fp}(p)\rangle=|\uparrow\uparrow...\uparrow\rangle$ is already an eigenstate of the system with the eigenvalue $\lambda_{fp}=1$.
Since cyclic permutation does not affect the configuration, we then have $N_j=1$ and only the $p=0$ condition is allowed.
The state with $S=N/2$ is $N+1$ times degenerated with respect to the different values of $S^z=-N/2,-N/2+1,...N/2$ and all
of those states have the only one available value of $p=0$.
The energy of the fully polarized system equals the total energy produced by the free spinless fermions, as expected.

\begin{figure}
\includegraphics[width=1\linewidth]{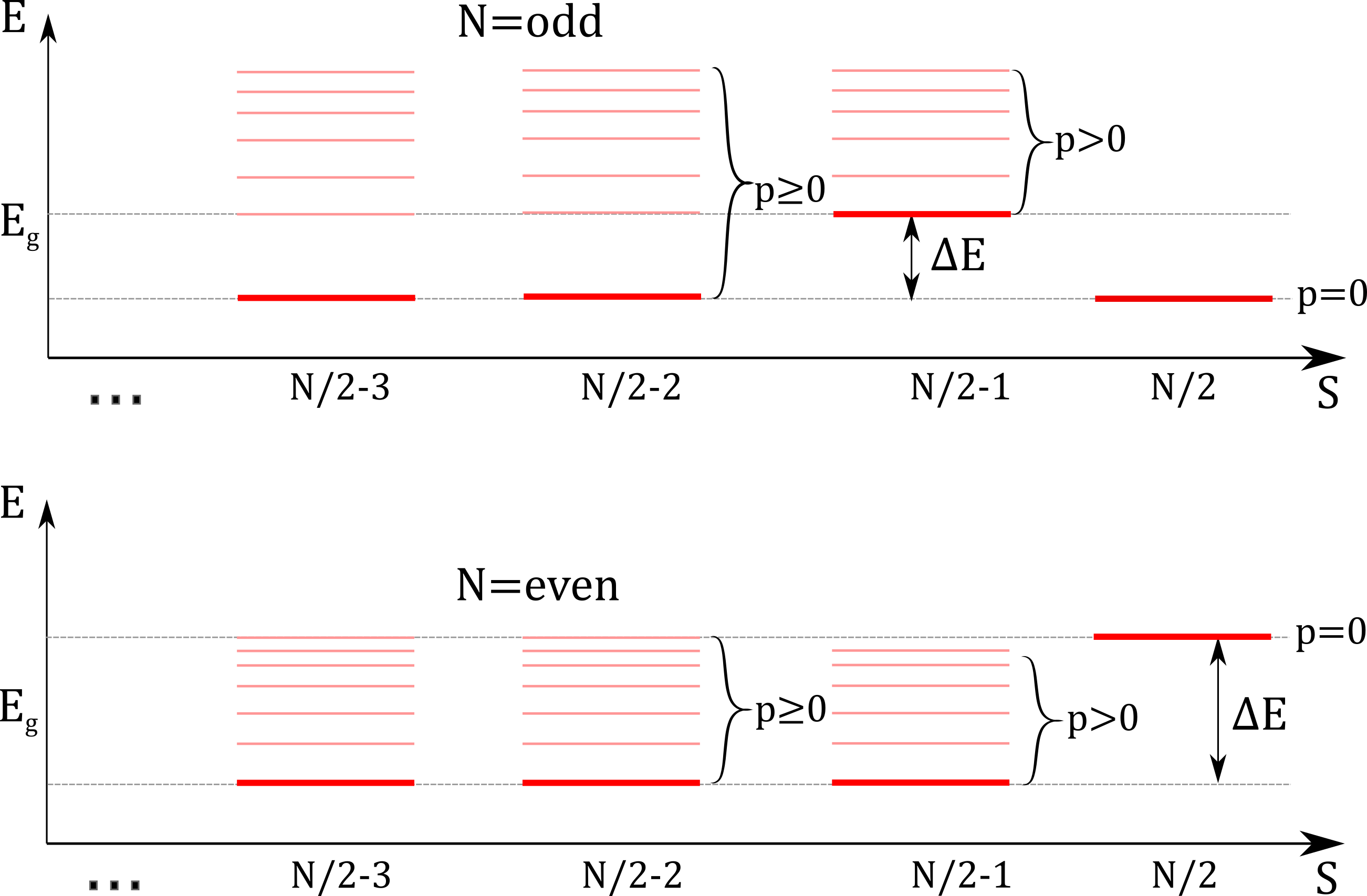}
\caption{The ground-state energy structure for different electron numbers $N$.}
\label{Fig2}
\end{figure}

Now let's consider a spin flipped case $|\psi_{sf}\rangle=|\downarrow\uparrow...\uparrow\rangle$.
There are $N$ states with $S^z=\frac{N}{2}-1$, now $N_j=N$ and all of these states are connected by cyclic permutations.
The eigenstates have the form:
\begin{equation}
|\psi_{sf}(p)\rangle=\frac{1}{\sqrt{N}}\sum_{q=0}^{N-1}e^{2\pi i pq/N}\cdot\hat{P}^{q}~|\downarrow\uparrow...\uparrow\rangle
\end{equation}
with the eigenvalues $\lambda_{sf}(p)=e^{2\pi i p/N}$ and the spin wave momentum runs over $p=0..N-1$.
Here the $p=0$ state is the $S=N/2$ state with the spin projection $S^z=N/2-1$ while the states with $p>0$ correspond to the total spin $S=N/2-1$ and $S^z=N/2-1$.
As a result, the fully polarized case $S=N/2$ with all possible projections $S^z$ has the only one available $p=0$ value for the
spin wave momentum while for the one spin flipped case $S=N/2-1$  the available values are $p>0$.
It is easy to see that for any value of the spin projection $S^z$ there must be one state with $S=N/2$ and $p=0$ and also $N-1$ states with $S=N/2-1$  and $p>0$.

Finally, for the states with $S^z\le N/2-2$ we always have more than one disconnected spin block,
which leads to the degeneracy of the levels.
As an example, in the case $S^z=N/2-2$ we have $\lfloor N/2\rfloor$ disconnected blocks and the $p=0$ level is $\lfloor N/2\rfloor$ fold degenerate.
While one of this degenerate levels is defined as the state with $S=N/2$ and $S^z=N/2-2$, all the other $\lfloor N/2\rfloor-1$ states correspond to the $S=N/2-2$ total spin (the $S=N/2-1$ states with $p=0$ being forbidden).
As a result both the $p=0$ and the $p>0$ values are available for the state with $S=N/2-2$.
Exactly the same behavior is a characteristic feature of the states with $S<N/2-2$.

Due to the fact that the exact eigenvalues in Eq.(\ref{eig}) are well known it is easy to analyse how the ground-state energy depends
explicitly on $p$.
In the case of an even number of electrons the presence of the spin wave with the nonzero  value of $p$ leads to decreasing of
the eigenenergy with the minimum reached at $p=N/2$.
Considering that the $p=N/2$ value is available for all values of spin except for $S=N/2$ the fully ferromagnetic state cannot be the ground state.
Due to the degeneracy of the $p=N/2$ state the ground state is a mixture of the states with the total spin $S\le N/2-1$.

In the case of an odd number of electrons a nonzero value of $p$ leads to the increase of the energy.
While the minimum of the energy corresponds to $p=0$, the next energy level corresponds to $p=1$.
As a result, the states with  $S=N/2$ and $S\le N/2-2$ have the same minimal value of the energy
with $p=0$ while the state with $S=N/2-1$ has only the $p=1$ level which is higher than the $p=0$ one.
The fully FM state with $S=N/2$ is separated from the states with $S\le N/2-2$ by an energy barrier $\Delta E$
due to the existence of the state $S=N/2-1$ with higher energy (Fig.\ref{Fig2}).
A one-spin flip process implies that the system goes over into the
higher energy state, which costs a certain energy.
Explicitly, a value of such \textit{a Nagaoka barrier} is
\begin{equation}
\Delta E=2t\sum_{k=-(N-1)/2}^{(N-1)/2}[\cos(\frac{2\pi}{L}k)-\cos(\frac{2\pi}{L}(k+\frac{1}{N}))].
\end{equation}

\begin{figure*}
\includegraphics[width=1\linewidth]{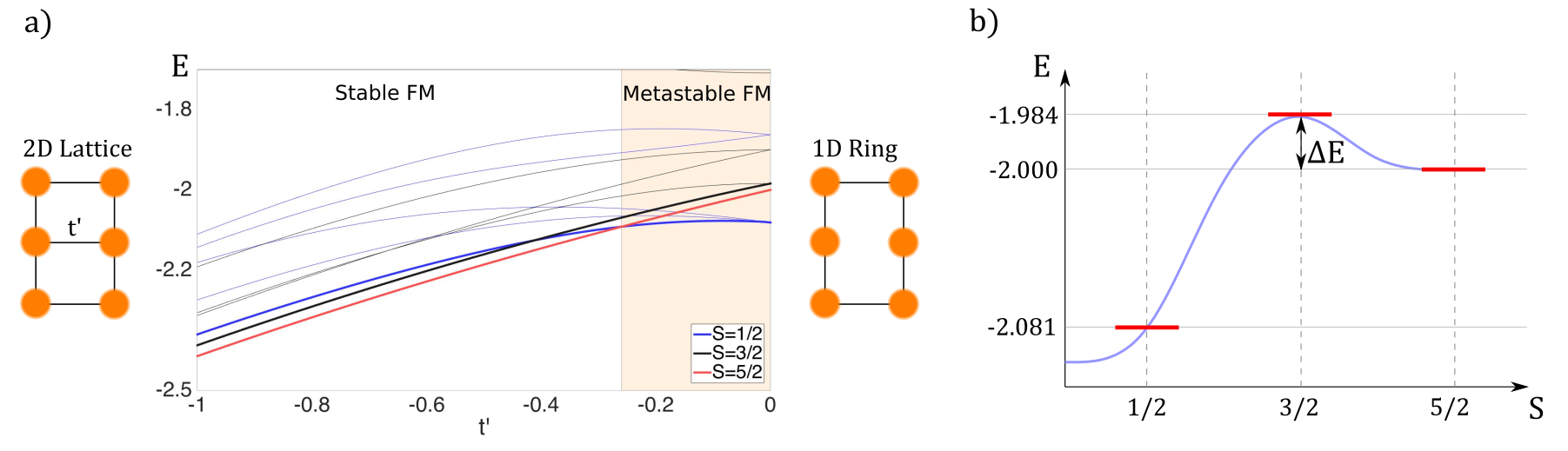}
\caption{\textbf{a)} The transition between stable and metastable Nagaoka ferromagnetism with a change in the topology of the quantum dot array from $2D$ to $1D$ (changing the $t'$ value).
\textbf{b)} The ground states energies for the different values of total spin $S$ for the $t-J$ model with $J=0.05t$, $L=6$, $N=5$.}
\label{Fig3}
\end{figure*}

The ground state of the odd number of electrons is thus a mixture of the FM $S=N/2$ state with the $S\le N/2-2$ states which are not true FM ones.
The only one possibility to get the true FM state is to put three electrons on a ring, in which case the states with $S\le N/2-2$ are absent.
There are only the $S=3/2$ and $S=1/2$ states available and
\begin{equation}
\Delta E=4t\sin^2(\frac{\pi}{3L})(1+2\cos(\frac{2\pi}{L})),
\end{equation}
with the condition that $L>3$.
Moreover, this barrier survives not only in the presence of infinity strong  on-site Coulomb repulsion but also
in more realistic cases including a weak enough short-range Coulolomb repulsion, antiferromagnetic and spin-orbit interactions, etc.\cite{Wang2019}.

In the case of odd $N>3$ the ground state is spin degenerate.
Realistic interactions remove this degeneracy.
Since the antiferromagnetic exchange interaction has a major impact on the system compared to the other possible interactions
the ground state tends to that of the minimum spin magnitude so that the global minimum in the energy corresponds to the non-ferromagnetic state.
On the other hand, for weak enough spin-spin exchange interaction the $S=N/2$ state remains separated from the $S\le N/2-2$ states by the barrier  rendering thereby the FM state a local minimum in the energy.

In the limit of $N\gg 1$ and $L \gg 1,\ N/L=n_e$- fixed
the $\Delta E$ reduces to
\begin{equation}
\Delta E=\frac{2tL}{\pi}\sin(\pi n_e)(1-\cos(\frac{2\pi}{NL}))\approx\frac{4\pi t}{N^2L}\sin( \pi n_e ).
\end{equation}
It is therefore clear that the Nagaoka barrier vanishes in the thermodynamic limit.
Both the presence of the \textit{disturbing} interactions and the rapid decrease of the barrier with the ring size and with the number of electrons could seem to hinder experimental observation of this phenomenon. However it can manifest itself in  quantum dot arrays through the appearance of the \textit{metastable} FM states.



To explicitly demonstrate how a metastable FM state can be experimentally realized,
let us consider the $2\times 3$ quantum dot array in the framework of $t-J$ model (Eq.\ref{Eq1}) with the large enough value of the Coulomb repulsion corresponding to $J=0.05$
and the hopping amplitude be $t'$ (Fig.\ref{Fig3}(a)) measured in units of $t$.
Varying $t'$ from $-1$ to $0$ corresponds to changing of the dimension of the array from the $2D$ cluster to the $1D$ ring.
Upon cooling down to a sufficiently low temperature the $2D$ cluster with $t'\neq 0$
displays the fully polarized FM ground state according to Nagaoka's theorem.
The breaking of the $t'$ link leads to the decrease of the $S=1/2$ state energy relative to the $S=3/2$ and $S=5/2$ states.
However, the $S=5/2$ state remains with lower energy than the $S=3/2$ one at all possible values of $t'$ thereby protecting electrons
from the spin flipping by the barrier $\Delta E$ even at a small but finite $J$.
If the temperature is low enough the five electron system remains in the metastable $S=5/2$ state for a finite time.
It should be noted that the system cooled at the $t'=0$ goes over to the global-minimum $S=1/2$ state.
In quantum dots experiments, the typical values of $t\approx 10^{2}\mu eV$ whereas the value of the barrier
is $\Delta E\approx10^{-2}t\approx 1\mu eV$.
To observe such an effect the condition $kT\ll\Delta E$ or $T\ll 10^{-2} K$ should necessarily be satisfied.
The fact that modern experiments are carried out already at temperatures of the order of $T\approx 10-100~mK$ allows us
to suggest that experimental observation of such an effect is likely to happen soon.
\paragraph*{Conclusion.}

It is shown that a problem of $N$ constrained spinfull electrons on a $L$-site ring reduces to that of the spinless electrons with
an effective flux threading through the ring. The flux arises due to the nonconnectivity of the emergent spin configurations.
It gives rise to the electron momentum quantization induced by the cyclic permutations group.
Nagaoka's ferromagnetism sets in  only in the case of three electrons on the ring with $L>3$.
For odd number of electrons greater than three the FM state can only be realized
as a local ground-state energy minimum protected by the Nagaoka barrier.
As a result, the fully polarized state for $N>3$ is truly a metastable state.
The progress in quantum dots fabrication provides a real possibility to observe such metastable ferromagnetism experimentally.
Specifically, we present experimental signatures of the metastable FM using a quantum dot device to host a $2\times 3$ lattice of electrons.
Effectively  tuning the geometry of the system from a $2D$ cluster to a $1D$ ring automatically drives the system into the metastable ground state.

\bibliography{Nagaoka_phase}

\begin{thebibliography}{14}%
\makeatletter
\providecommand \@ifxundefined [1]{%
 \@ifx{#1\undefined}
}%
\providecommand \@ifnum [1]{%
 \ifnum #1\expandafter \@firstoftwo
 \else \expandafter \@secondoftwo
 \fi
}%
\providecommand \@ifx [1]{%
 \ifx #1\expandafter \@firstoftwo
 \else \expandafter \@secondoftwo
 \fi
}%
\providecommand \natexlab [1]{#1}%
\providecommand \enquote  [1]{``#1''}%
\providecommand \bibnamefont  [1]{#1}%
\providecommand \bibfnamefont [1]{#1}%
\providecommand \citenamefont [1]{#1}%
\providecommand \href@noop [0]{\@secondoftwo}%
\providecommand \href [0]{\begingroup \@sanitize@url \@href}%
\providecommand \@href[1]{\@@startlink{#1}\@@href}%
\providecommand \@@href[1]{\endgroup#1\@@endlink}%
\providecommand \@sanitize@url [0]{\catcode `\\12\catcode `\$12\catcode
  `\&12\catcode `\#12\catcode `\^12\catcode `\_12\catcode `\%12\relax}%
\providecommand \@@startlink[1]{}%
\providecommand \@@endlink[0]{}%
\providecommand \url  [0]{\begingroup\@sanitize@url \@url }%
\providecommand \@url [1]{\endgroup\@href {#1}{\urlprefix }}%
\providecommand \urlprefix  [0]{URL }%
\providecommand \Eprint [0]{\href }%
\providecommand \doibase [0]{http://dx.doi.org/}%
\providecommand \selectlanguage [0]{\@gobble}%
\providecommand \bibinfo  [0]{\@secondoftwo}%
\providecommand \bibfield  [0]{\@secondoftwo}%
\providecommand \translation [1]{[#1]}%
\providecommand \BibitemOpen [0]{}%
\providecommand \bibitemStop [0]{}%
\providecommand \bibitemNoStop [0]{.\EOS\space}%
\providecommand \EOS [0]{\spacefactor3000\relax}%
\providecommand \BibitemShut  [1]{\csname bibitem#1\endcsname}%
\let\auto@bib@innerbib\@empty
\bibitem [{\citenamefont {Dehollain}\ \emph {et~al.}()\citenamefont
  {Dehollain}, \citenamefont {Mukhopadhyay}, \citenamefont {Michal},
  \citenamefont {Wang}, \citenamefont {Wunsch}, \citenamefont {Reichl},
  \citenamefont {Wegscheider}, \citenamefont {Rudner}, \citenamefont {Demler},\
  and\ \citenamefont {Vandersypen}}]{Dehollain2019}%
  \BibitemOpen
  \bibfield  {author} {\bibinfo {author} {\bibfnamefont {J.~P.}\ \bibnamefont
  {Dehollain}}, \bibinfo {author} {\bibfnamefont {U.}~\bibnamefont
  {Mukhopadhyay}}, \bibinfo {author} {\bibfnamefont {V.~P.}\ \bibnamefont
  {Michal}}, \bibinfo {author} {\bibfnamefont {Y.}~\bibnamefont {Wang}},
  \bibinfo {author} {\bibfnamefont {B.}~\bibnamefont {Wunsch}}, \bibinfo
  {author} {\bibfnamefont {C.}~\bibnamefont {Reichl}}, \bibinfo {author}
  {\bibfnamefont {W.}~\bibnamefont {Wegscheider}}, \bibinfo {author}
  {\bibfnamefont {M.~S.}\ \bibnamefont {Rudner}}, \bibinfo {author}
  {\bibfnamefont {E.}~\bibnamefont {Demler}}, \ and\ \bibinfo {author}
  {\bibfnamefont {L.~M.~K.}\ \bibnamefont {Vandersypen}},\ }\href@noop {} {\
  }\Eprint {http://arxiv.org/abs/http://arxiv.org/abs/1904.05680v2}
  {http://arxiv.org/abs/1904.05680v2} \BibitemShut {NoStop}%
\bibitem [{\citenamefont {Takahashi}(1982)}]{Takahashi1982}%
  \BibitemOpen
  \bibfield  {author} {\bibinfo {author} {\bibfnamefont {M.}~\bibnamefont
  {Takahashi}},\ }\href {\doibase 10.1143/jpsj.51.3475} {\bibfield  {journal}
  {\bibinfo  {journal} {Journal of the Physical Society of Japan}\ }\textbf
  {\bibinfo {volume} {51}},\ \bibinfo {pages} {3475} (\bibinfo {year}
  {1982})}\BibitemShut {NoStop}%
\bibitem [{\citenamefont {Buterakos}\ and\ \citenamefont
  {Sarma}()}]{Buterakos2019}%
  \BibitemOpen
  \bibfield  {author} {\bibinfo {author} {\bibfnamefont {D.}~\bibnamefont
  {Buterakos}}\ and\ \bibinfo {author} {\bibfnamefont {S.~D.}\ \bibnamefont
  {Sarma}},\ }\href@noop {} {\ }\Eprint
  {http://arxiv.org/abs/http://arxiv.org/abs/1908.03226v1}
  {http://arxiv.org/abs/1908.03226v1} \BibitemShut {NoStop}%
\bibitem [{\citenamefont {Wang}\ \emph {et~al.}()\citenamefont {Wang},
  \citenamefont {Dehollain}, \citenamefont {Liu}, \citenamefont {Mukhopadhyay},
  \citenamefont {Rudner}, \citenamefont {Vandersypen},\ and\ \citenamefont
  {Demler}}]{Wang2019}%
  \BibitemOpen
  \bibfield  {author} {\bibinfo {author} {\bibfnamefont {Y.}~\bibnamefont
  {Wang}}, \bibinfo {author} {\bibfnamefont {J.~P.}\ \bibnamefont {Dehollain}},
  \bibinfo {author} {\bibfnamefont {F.}~\bibnamefont {Liu}}, \bibinfo {author}
  {\bibfnamefont {U.}~\bibnamefont {Mukhopadhyay}}, \bibinfo {author}
  {\bibfnamefont {M.~S.}\ \bibnamefont {Rudner}}, \bibinfo {author}
  {\bibfnamefont {L.~M.~K.}\ \bibnamefont {Vandersypen}}, \ and\ \bibinfo
  {author} {\bibfnamefont {E.}~\bibnamefont {Demler}},\ }\href {\doibase
  10.1103/PhysRevB.100.155133} {\ 10.1103/PhysRevB.100.155133},\ \Eprint
  {http://arxiv.org/abs/http://arxiv.org/abs/1907.01658v3}
  {http://arxiv.org/abs/1907.01658v3} \BibitemShut {NoStop}%
\bibitem [{\citenamefont {Nagaoka}(1965)}]{Nagaoka1965}%
  \BibitemOpen
  \bibfield  {author} {\bibinfo {author} {\bibfnamefont {Y.}~\bibnamefont
  {Nagaoka}},\ }\href {\doibase 10.1016/0038-1098(65)90266-8} {\bibfield
  {journal} {\bibinfo  {journal} {Solid State Communications}\ }\textbf
  {\bibinfo {volume} {3}},\ \bibinfo {pages} {409} (\bibinfo {year}
  {1965})}\BibitemShut {NoStop}%
\bibitem [{\citenamefont {Nagaoka}(1966)}]{Nagaoka1966}%
  \BibitemOpen
  \bibfield  {author} {\bibinfo {author} {\bibfnamefont {Y.}~\bibnamefont
  {Nagaoka}},\ }\href {\doibase 10.1103/physrev.147.392} {\bibfield  {journal}
  {\bibinfo  {journal} {Physical Review}\ }\textbf {\bibinfo {volume} {147}},\
  \bibinfo {pages} {392} (\bibinfo {year} {1966})}\BibitemShut {NoStop}%
\bibitem [{\citenamefont {Becca}\ and\ \citenamefont
  {Sorella}(2001)}]{Becca2001}%
  \BibitemOpen
  \bibfield  {author} {\bibinfo {author} {\bibfnamefont {F.}~\bibnamefont
  {Becca}}\ and\ \bibinfo {author} {\bibfnamefont {S.}~\bibnamefont
  {Sorella}},\ }\href {\doibase 10.1103/physrevlett.86.3396} {\bibfield
  {journal} {\bibinfo  {journal} {Physical Review Letters}\ }\textbf {\bibinfo
  {volume} {86}},\ \bibinfo {pages} {3396} (\bibinfo {year}
  {2001})}\BibitemShut {NoStop}%
\bibitem [{\citenamefont {Liu}\ \emph {et~al.}(2012)\citenamefont {Liu},
  \citenamefont {Yao}, \citenamefont {Berg}, \citenamefont {White},\ and\
  \citenamefont {Kivelson}}]{Liu2012}%
  \BibitemOpen
  \bibfield  {author} {\bibinfo {author} {\bibfnamefont {L.}~\bibnamefont
  {Liu}}, \bibinfo {author} {\bibfnamefont {H.}~\bibnamefont {Yao}}, \bibinfo
  {author} {\bibfnamefont {E.}~\bibnamefont {Berg}}, \bibinfo {author}
  {\bibfnamefont {S.~R.}\ \bibnamefont {White}}, \ and\ \bibinfo {author}
  {\bibfnamefont {S.~A.}\ \bibnamefont {Kivelson}},\ }\href {\doibase
  10.1103/physrevlett.108.126406} {\bibfield  {journal} {\bibinfo  {journal}
  {Physical Review Letters}\ }\textbf {\bibinfo {volume} {108}} (\bibinfo
  {year} {2012}),\ 10.1103/physrevlett.108.126406}\BibitemShut {NoStop}%
\bibitem [{\citenamefont {Putikka}\ \emph {et~al.}(1992)\citenamefont
  {Putikka}, \citenamefont {Luchini},\ and\ \citenamefont
  {Ogata}}]{Putikka1992}%
  \BibitemOpen
  \bibfield  {author} {\bibinfo {author} {\bibfnamefont {W.~O.}\ \bibnamefont
  {Putikka}}, \bibinfo {author} {\bibfnamefont {M.~U.}\ \bibnamefont
  {Luchini}}, \ and\ \bibinfo {author} {\bibfnamefont {M.}~\bibnamefont
  {Ogata}},\ }\href {\doibase 10.1103/physrevlett.69.2288} {\bibfield
  {journal} {\bibinfo  {journal} {Physical Review Letters}\ }\textbf {\bibinfo
  {volume} {69}},\ \bibinfo {pages} {2288} (\bibinfo {year}
  {1992})}\BibitemShut {NoStop}%
\bibitem [{\citenamefont {Ivantsov}\ \emph {et~al.}(2017)\citenamefont
  {Ivantsov}, \citenamefont {Ferraz},\ and\ \citenamefont
  {Kochetov}}]{Ivantsov2017}%
  \BibitemOpen
  \bibfield  {author} {\bibinfo {author} {\bibfnamefont {I.}~\bibnamefont
  {Ivantsov}}, \bibinfo {author} {\bibfnamefont {A.}~\bibnamefont {Ferraz}}, \
  and\ \bibinfo {author} {\bibfnamefont {E.}~\bibnamefont {Kochetov}},\ }\href
  {\doibase 10.1103/physrevb.95.155115} {\bibfield  {journal} {\bibinfo
  {journal} {Physical Review B}\ }\textbf {\bibinfo {volume} {95}} (\bibinfo
  {year} {2017}),\ 10.1103/physrevb.95.155115}\BibitemShut {NoStop}%
\bibitem [{\citenamefont {Kornilovitch}(2014)}]{Kornilovitch2014}%
  \BibitemOpen
  \bibfield  {author} {\bibinfo {author} {\bibfnamefont {P.}~\bibnamefont
  {Kornilovitch}},\ }\href {\doibase 10.1103/physrevlett.112.077202} {\bibfield
   {journal} {\bibinfo  {journal} {Physical Review Letters}\ }\textbf {\bibinfo
  {volume} {112}} (\bibinfo {year} {2014}),\
  10.1103/physrevlett.112.077202}\BibitemShut {NoStop}%
\bibitem [{\citenamefont {Lieb}\ and\ \citenamefont {Mattis}(1962)}]{Lieb1962}%
  \BibitemOpen
  \bibfield  {author} {\bibinfo {author} {\bibfnamefont {E.}~\bibnamefont
  {Lieb}}\ and\ \bibinfo {author} {\bibfnamefont {D.}~\bibnamefont {Mattis}},\
  }\href {\doibase 10.1103/physrev.125.164} {\bibfield  {journal} {\bibinfo
  {journal} {Physical Review}\ }\textbf {\bibinfo {volume} {125}},\ \bibinfo
  {pages} {164} (\bibinfo {year} {1962})}\BibitemShut {NoStop}%
\bibitem [{\citenamefont {Xavier}\ \emph {et~al.}()\citenamefont {Xavier},
  \citenamefont {Ferraz},\ and\ \citenamefont {Kochetov}}]{Xavier2019}%
  \BibitemOpen
  \bibfield  {author} {\bibinfo {author} {\bibfnamefont {H.~B.}\ \bibnamefont
  {Xavier}}, \bibinfo {author} {\bibfnamefont {A.}~\bibnamefont {Ferraz}}, \
  and\ \bibinfo {author} {\bibfnamefont {E.}~\bibnamefont {Kochetov}},\
  }\href@noop {} {\ }\Eprint
  {http://arxiv.org/abs/http://arxiv.org/abs/1909.07349v2}
  {http://arxiv.org/abs/1909.07349v2} \BibitemShut {NoStop}%
\bibitem [{\citenamefont {Haerter}\ and\ \citenamefont
  {Shastry}(2005)}]{Haerter2005}%
  \BibitemOpen
  \bibfield  {author} {\bibinfo {author} {\bibfnamefont {J.~O.}\ \bibnamefont
  {Haerter}}\ and\ \bibinfo {author} {\bibfnamefont {B.~S.}\ \bibnamefont
  {Shastry}},\ }\href {\doibase 10.1103/physrevlett.95.087202} {\bibfield
  {journal} {\bibinfo  {journal} {Physical Review Letters}\ }\textbf {\bibinfo
  {volume} {95}} (\bibinfo {year} {2005}),\
  10.1103/physrevlett.95.087202}\BibitemShut {NoStop}%
\end{thebibliography}%
\end{document}